\documentclass[aps,showpacs,preprintnumbers,nofootinbibt, preprint]{revtex4}
\usepackage{graphicx}

\begin{document}

\title{Bose-Einstein condensation of dark matter solves the core/cusp problem}
\author{T. Harko}
\email{harko@hkucc.hku.hk}
\affiliation{Department of Physics and
Center for Theoretical and Computational Physics, The University
of Hong Kong, Pok Fu Lam Road, Hong Kong, P. R. China}

\begin{abstract}
We analyze the observed properties of dwarf galaxies, which are dark matter dominated astrophysical objects, by assuming that dark matter is in the form of a strongly - coupled, dilute Bose - Einstein condensate. The basic astrophysical properties of the condensate (density profile, rotational velocity, and mass profile, respectively), are derived from a variational principle.  To test the
validity of the model we compare first the tangential velocity equation of the model with a sample
of eight rotation curves of dwarf galaxies. We find a  good agreement between the theoretically  predicted rotation curves (without any baryonic component) and the observational data. The mean value of the logarithmic inner slope of the mass density profile of dwarf galaxies is also obtained, and it is shown that the observed value of this parameter is in agreement with the theoretical results. The predictions of the Bose - Einstein condensate model are also systematically compared with the predictions of the standard Cold Dark Matter model. The non-singular density profiles of the Bose-Einstein condensed dark matter generally show the presence of an extended core, whose presence is due to the strong interaction between dark matter particles.
\end{abstract}


\date{\today}

\maketitle

\section{Introduction}

Despite the important achievements of the  $\Lambda $ Cold Dark Matter ($\Lambda $CDM)  model \cite{PeRa03},
at galactic scales $\sim 10$ kpc, the standard cosmological model meets with
severe difficulties in explaining the observed
distribution of the invisible matter around the luminous one. In fact, $N$%
-body simulations, performed in this scenario, predict that bound halos
surrounding galaxies must have very characteristic density profiles, called the Navarro-Frenk-White (NFW) profiles, which feature a well pronounced central cusp \cite{nfw},
\begin{equation}\label{NFW}
\rho _{NFW}(r)=\frac{\rho
_{s}}{(r/r_{s})(1+r/r_{s})^{2}} ,
\end{equation}
where $r_{s}$ is a scale radius
and $\rho _{s}$ is a characteristic density. On the observational side,
high-resolution rotation curves show, instead, that the actual distribution
of dark matter is much shallower than the above, and it presents a nearly
constant density core \cite{bur}.

The discrepancies between observations and simulations done in the framework
of the $\Lambda $CDM model are particularly important in the case of the recent
observations of the dwarf galaxies done by ''The Hi Nearby Galaxy Survey''
(THINGS) \cite{Wa}. Dwarf galaxies are dark matter dominated cosmic
structures, with a very small contribution of baryons to the total matter
content. Similarly to the Low Surface Brightness (LSB) galaxies, they are
ideal objects for the study of the dark matter properties. The observed
rotation curves of the dwarf galaxies rise too slowly, as compared to the
rotation curves derived from the cusp-like dark matter distribution in CDM
halos. Instead, they are better described by core-like models, dominated by a
central constant-density core \cite{Oh}. Moreover, the mean value of the
logarithmic inner slopes of the mass density profiles of the dwarf galaxies
is $\alpha =-0.29\pm 0.07$ \cite{Oh}, while for a large sample of LSB
galaxies $\alpha =-0.2\pm 0.2$ \cite{deBl}, values which are significantly
different from the steep slope of $\alpha =-1.0$, inferred from
dark-matter-only simulations. A second serious problem is related to the
mass of the dwarf galaxies obtained from simulations. The dwarf galaxies
formed in current hydrodynamical simulations are more than an order of
magnitude more luminous than expected for haloes of masses of the order of $%
10^{10}M_{\odot}$ \cite{Ti}. Therefore explaining the observed properties of dwarf
galaxies has become a very challenging task, and ''missing astrophysical
effects in the simulations are the most likely cause of the discrepancy, and
the most promising target in search of its resolution'' \cite{Ti}.

Recently, it was suggested that cold dark matter
particles interacting through a Yukawa potential could produce the cores in
dwarf galaxies \cite{Lo}. Simulations that include the effect of baryonic
feedback processes, such as gas cooling, star formation, cosmic UV
background heating and  physically motivated gas outflows
driven by supernovae have succeeded in obtaining a value of $\alpha =-0.4\pm
0.1$ \cite{Oh1}.

The possibility that dark matter is in the form of a Bose-Einstein Condensate (BEC), namely, an assembly of light individual bosons that acquire a repulsive interaction by occupying the same ground energy state, has been  extensively investigated in the physical literature \cite{Sin, fer, Fu05,fer1, BoHa07}. The experimental realization of the Bose-Einstein condensation in terrestrial laboratories for a large class of particles (atoms) and physical systems  has given a sound theoretical and experimental support to this hypothesis.

The Bose-Einstein condensation process was first observed experimentally in
1995 in dilute alkali gases, such as vapors of rubidium and sodium, confined
in a magnetic trap, and cooled to very low temperatures. A sharp peak in the
velocity distribution was observed below a critical temperature, indicating
that condensation has occurred, with the alkali atoms condensed in the same
ground state and showing a narrow peak in the momentum space and in the
coordinate space \cite{exp}.

Different properties of the Bose-Einstein condensed dark matter were investigated recently in \cite{bec1} - \cite{Guz}. It was also shown
that the creation of quantum vortices are favored in
strongly-coupled condensates, while this is not the case for axion
condensates, in which the particles are effectively non-interacting \cite{Ri}.

The possibility that the core-cusp problem of standard CDM models can be solved by assuming that  dark matter is
composed of ultralight scalar particles, with masses of the order of $10^{-22}$ eV, initially in a cold Bose-Einstein condensate ( "fuzzy dark matter"), was considered in \cite{Hu}.
 The wave properties of the dark matter stabilize gravitational
collapse, providing halo cores. In this model stability below the Jeans scale is guaranteed by the uncertainty principle, since an increase in momentum opposes any attempt to confine
the particle further. The
one-dimensional simulations performed in \cite{Hu} suggest that the density profile on
the Jeans scale  is not universal, but also evolves continuously
on the dynamical time scale (or faster), due to
quantum interference effects.

An analysis of the rotation curves in the Gross-Pitaevskii equation based BEC model of dark matter was performed, for several High Surface Brightness (HSB), LSB, and dwarf galaxies, respectively, in \cite{BoHa07}. There is a good general agreement between the theoretical predictions and observations. However, there are several reasons to extend the analysis initiated in \cite{BoHa07}. First, there is a significant increase in the observational accuracy of rotation curve data for dwarf galaxies, which are dark matter dominated objects, and more high precision data are available presently \cite{Wa,Oh}. Secondly, there is a major discrepancy between observations and the theoretical predictions of the CDM models for the  mean value
of the logarithmic inner density slopes of the dwarf galaxies. The mean values
of the logarithmic inner density and velocity slopes are observed quantities that could prove more useful in discriminating between different dark matter models and observations than the rotation curves themselves, which can be fitted by many models. Therefore it is important to analyze these quantities in the framework of the BEC models. Thirdly, the explicit presentation and detailed discussion of the BEC density profiles for a sample of dwarf galaxies has not been done yet. And finally, a systematic comparison between the prediction of the Navarro - Frenk - White  dark matter density profiles, given by Eq.~(\ref{NFW}), and the BEC profiles, is still missing.

It is the main goal of the present paper to address the above mentioned issues, by analyzing the observed properties
of the dark matter halos of eight dwarf galaxies. The dark matter is assumed to be in the form of a strongly - coupled, non-relativistic diluted Bose-Einstein condensate, as proposed in \cite{BoHa07}.  As a first step in our study we derive the basic astrophysical parameters of the condensate (density, velocity and mass profiles, respectively) from a variational principle. Then the theoretical predictions of the model are compared with the observational data for eight dwarf galaxies.  The theoretical predictions of the rotation curves are compared with  the observed rotation curves of the dwarf galaxies, as well as with the results derived from the NFW profile.  There is a general good agreement between the theoretical predictions of the BEC model and the behavior of the rotation curves. The observed mean value of the logarithmic inner slope of the mass density profile can also be theoretically reproduced. The dark matter density profiles for these galaxies are also explicitly presented, and compared with the standard NFW profiles.

The present paper is organized as follows. In Section~\ref{2} the basic equations of the BEC model are written down. The predictions of the BEC and NFW models are compared with the observations in Section~\ref{3}. We discuss and conclude our results in Section~\ref{4}.

\section{Bose-Einstein condensed dark matter}\label{2}

At very low temperatures, all particles in a dilute Bose gas condense to the
same quantum ground state, forming a Bose-Einstein Condensate (BEC). Particles become correlated with each other when their wavelengths
overlap, that is, the thermal wavelength $\lambda _{T}$ is greater than the
mean inter-particles distance $l$. This happens at a temperature $T<2\pi
\hbar ^{2}/mk_{B}n^{2/3}$, where $m$ is the mass of the particle in the
condensate, $n$ is the number density, and $k_{B}$ is Boltzmann's constant
\citep{Da99}. A coherent state develops when the particle density is enough
high, or the temperature is sufficiently low.
We assume that the dark matter halos are composed of a strongly - coupled dilute Bose-Einstein
condensate at absolute zero. Hence almost all the dark matter particles are
in the condensate.  In a dilute and cold gas,
only binary collisions at low energy are relevant, and these collisions are
characterized by a single parameter, the $s$-wave scattering length $a$,
independently of the details of the two-body potential. Therefore, one can
replace the interaction potential with an effective
interaction $V_I\left( \vec{r}^{\prime }-\vec{r}\right) =\lambda \delta \left(
\vec{r}^{\prime }-\vec{r}\right) $, where the coupling constant $\lambda $
is related to the scattering length $a$
through $\lambda =4\pi \hbar ^{2}a/m$ \cite{Da99}.
The ground state properties of the dark matter are
described by the mean-field Gross-Pitaevskii (GP) equation. The GP equation
for the dark matter halos can be derived from the GP energy functional,
\begin{eqnarray}
E\left[ \psi \right] &=&\int \left[ \frac{\hbar ^{2}}{2m}\left| \nabla \psi
\left( \vec{r}\right) \right| ^{2}+\frac{U_{0}}{2}\left| \psi \left( \vec{r}%
\right) \right| ^{4}\right] d\vec{r}-\nonumber\\
&&\frac{1}{2}Gm^{2}\int \int \frac{\left|
\psi \left( \vec{r}\right) \right| ^{2}\left| \psi \left( \vec{r}^{\prime
}\right) \right| ^{2}}{\left| \vec{r}-\vec{r}^{\prime }\right| }d\vec{r}d%
\vec{r}^{\prime },
\end{eqnarray}
where $\psi \left( \vec{r}\right) $ is the wave function of the condensate,
and $U_{0}=4\pi \hbar ^{2}a/m$ \cite{Da99}. The first term in the energy functional is
the quantum pressure, the second is the interaction energy, and the third is
the gravitational potential energy. The mass density of the condensate is
defined as $\rho \left( \vec{r}\right) =m\left| \psi \left( \vec{r}\right)
\right| ^{2}$, and the normalization condition is $N=\int \left| \psi \left(
\vec{r}\right) \right| ^{2}d\vec{r}$, where $N$ is the total number of dark
matter particles. The variational
procedure  $\delta E\left[ \psi \right] -\mu \delta \int \left| \psi \left(
\vec{r}\right) \right| ^{2}d\vec{r}=0$ gives the GP equation as
\begin{equation}
-\frac{\hbar ^{2}}{2m}\nabla ^{2}\psi \left( \vec{r}\right) +V\left( \vec{r}%
\right) \psi \left( \vec{r}\right) +U_{0}\left| \psi \left( \vec{r}\right)
\right| ^{2}\psi \left( \vec{r}\right) =\mu \psi \left( \vec{r}\right) ,
\end{equation}
where $\mu $ is the chemical potential, and the gravitational potential $V$
satisfies the Poisson equation $\nabla ^{2}V=4\pi G\rho $. When the number
of particles in the condensate becomes large enough, the quantum pressure
term makes a significant contribution only near the boundary, and is much
smaller than the interaction energy term. Thus the quantum pressure term can
be neglected (the Thomas-Fermi approximation). When $N\rightarrow \infty $,
the Thomas-Fermi approximation becomes exact \cite{Da99}.  Therefore we obtain
\begin{equation}
\rho \left( \vec{r}\right) =\frac{m}{U_{0}} \left[ \mu
-mV\left( \vec{r}\right) \right] .
\end{equation}
 The Poisson equation becomes
 \begin{equation}
 \nabla ^{2}\rho +k^{2}\rho =0,
\end{equation}
where $k=\sqrt{4\pi Gm^{2}/U_{0}}$.

Hence the density
distribution $\rho $ of the static gravitationally bounded single
component dark matter Bose-Einstein condensate is given by  \cite{BoHa07}%
\begin{equation}
\rho \left( r\right) =\rho _{c}\frac{\sin kr}{kr}
\end{equation}
where  $\rho _{c}$ is the central
density of the condensate, $\rho _{c}=\rho (0)$. The density profile is non-singular at the center $r=0$. The mass
profile $m(r)=4\pi \int_{0}^{r}\rho (r)r^{2}dr$ of the
galactic dark matter halo is
\begin{equation}
m\left( r\right) =\frac{4\pi \rho r}{k^2}\left( 1-kr\cot kr\right),
\end{equation}
with a boundary radius $R$. At the boundary of the dark matter
distribution $\rho (R)=0$, giving the condition $kR=\pi $,
which fixes the radius of the condensate dark matter halo as
\begin{equation}
R=\pi
\sqrt{\frac{\hbar ^{2}a}{Gm^{3}}}.
\end{equation}
The tangential velocity $V$ of a test particle
moving in the condensed dark halo can be represented as
\begin{equation}\label{vel}
V^{2}\left(
r\right) =\frac{Gm(r)}{r}= \frac{4\pi G\rho }{k^{2}} \left(
1-kr\cot kr\right),
\end{equation}
or, alternatively,
\begin{equation}
V^2\left(r\right)=\frac{4G\rho _c R^2}{\pi }\left[\frac{\sin\left(\pi r/R \right)}{\pi r/R}-\cos\left(\frac{\pi r}{R}\right)\right].
\end{equation}
The total mass of the condensate dark matter halo $M$ can be obtained as
\begin{equation}\label{mass}
M=4\pi ^2\left(\frac{\hbar ^2a}{Gm^3}\right)^{3/2}\rho _c=\frac{4}{\pi }R^3\rho _c,
\end{equation}
giving for the mean value $<\rho >$ of the condensate density $<\rho >=3\rho _c/\pi ^2$. Hence the tangential velocity can also be represented as
\begin{equation}
V^2\left(r\right)=\frac{GM}{R}\left[\frac{\sin\left(\pi r/R \right)}{\pi r/R}-\cos\left(\frac{\pi r}{R}\right)\right].
\end{equation}
For $r>R$ the rotation curves follow the standard Keplerian law.
The mass of the particle in the
condensate is given by
\begin{equation}
 m =\left( \frac{\pi ^{2}\hbar ^{2}a}{GR^{2}}\right) ^{1/3}\approx
6.73\times 10^{-2}\left[ a\left( {\rm fm}\right) \right] ^{1/3}%
\left[ R\;{\rm (kpc)}\right] ^{-2/3}\;{\rm eV}.
\end{equation}
For $a\approx
1 $ fm and $R\approx 10$ kpc, the typical mass of the condensate particle is of the order of $m\approx 14$
meV. For $a\approx 10^{6}$ fm, corresponding
to the values of $a$ observed in terrestrial laboratory experiments, $%
m\approx 1.44$ eV. These values are consistent with the limit $%
m<1.87$ eV obtained for the mass of the condensate particle from
cosmological considerations \cite{Bo}.

\section{Properties of dwarf galaxies in the BEC model}\label{3}

In order to analyze the properties of the dwarf galaxies in the framework of the BEC and CDM models we have selected a sample of eight dwarf galaxies for which high resolution rotation curves are available. The galaxies we consider are IC 2574, NGC 2366, Holmberg I (Ho I),
Holmberg II (Ho II), M81 dwB, DDO 53 \cite{Oh}, DDO 39 \cite{Sw} \ and DDO
154 \cite{Ca}, respectively.  As a first step in the comparison of the observed properties of
dwarf galaxies with the BEC model and the CDM model, respectively,  we analyze the behavior of the rotation
curves, by comparing the predictions of Eq.~(\ref{vel}) and of Eq.~(\ref{NFW}) with the observed
rotation curves of the eight galaxies. This comparison allows us to determine either theoretically (BEC model), or by fitting (NFW model) the free parameters in the equations. Then, as a second step in our study we obtain the dark matter density profiles of the galaxies, and we estimate the mean values of the logarithmic density and velocity slopes in the BEC model.

\subsection{Rotational velocity profiles of the dwarf galaxies}

We begin our study of the BEC dark matter halo model by comparing  its theoretical predictions with the observational data of the rotation curves of eight dwarf galaxies, as well as with the velocity curves obtained by using the standard NFW profile, given by Eq.~(\ref{NFW}). For the NFW density profile the rotational velocity is given by
\begin{equation}
V_{NFW}(r)=\sqrt{4\pi G\rho _{s}r_{s}^{3}}\sqrt{\frac{1}{r}\left[ \ln \left(
1+\frac{r}{r_{s}}\right) -\frac{\left( r/r_{s}\right) }{1+r/r_{s}}\right] }.
\end{equation}

In order to obtain the rotation curves in the BEC model we use the following procedure. The BEC theoretical rotation curves {\it are not obtained by fitting the observational data}. Instead, a reasonable value for the radius $R$, interpreted as the point where the rotation curve starts to decline, is obtained from observations, and then a central density value is assumed. Therefore {\it the theoretical rotation curves are predictions of the model}, and not fits. On the other hand, we {\it fit} the NFW profiles with the observational data.

\begin{figure}[!h]
\centering
\includegraphics[width=0.48\linewidth]{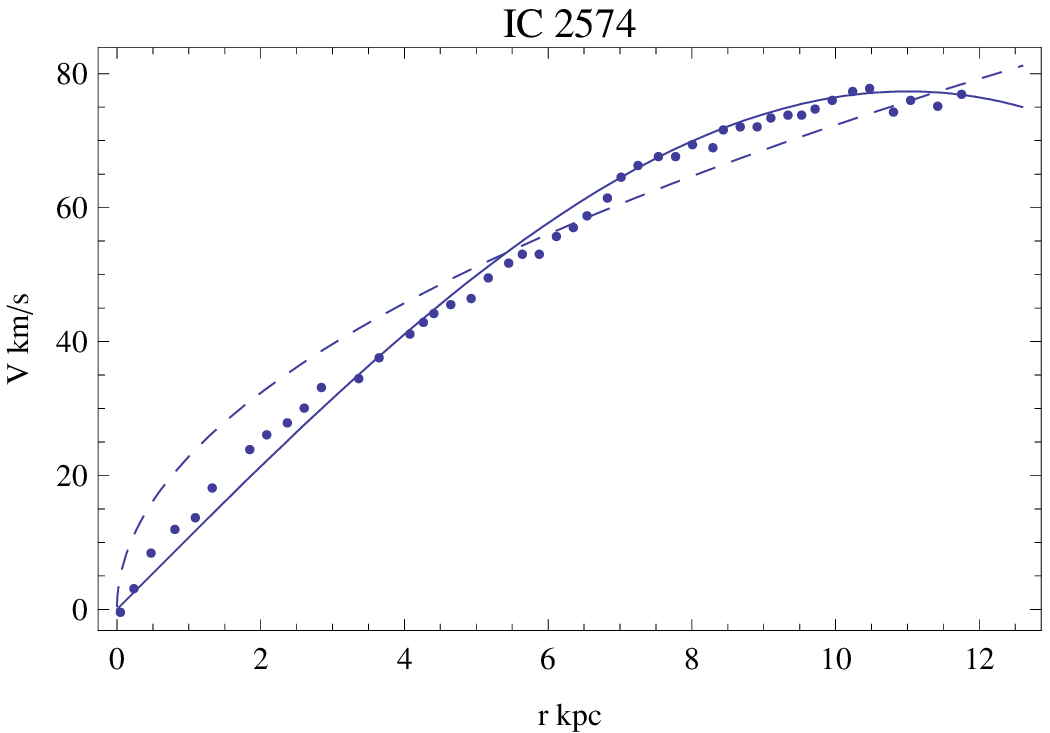}
\includegraphics[width=0.48\linewidth]{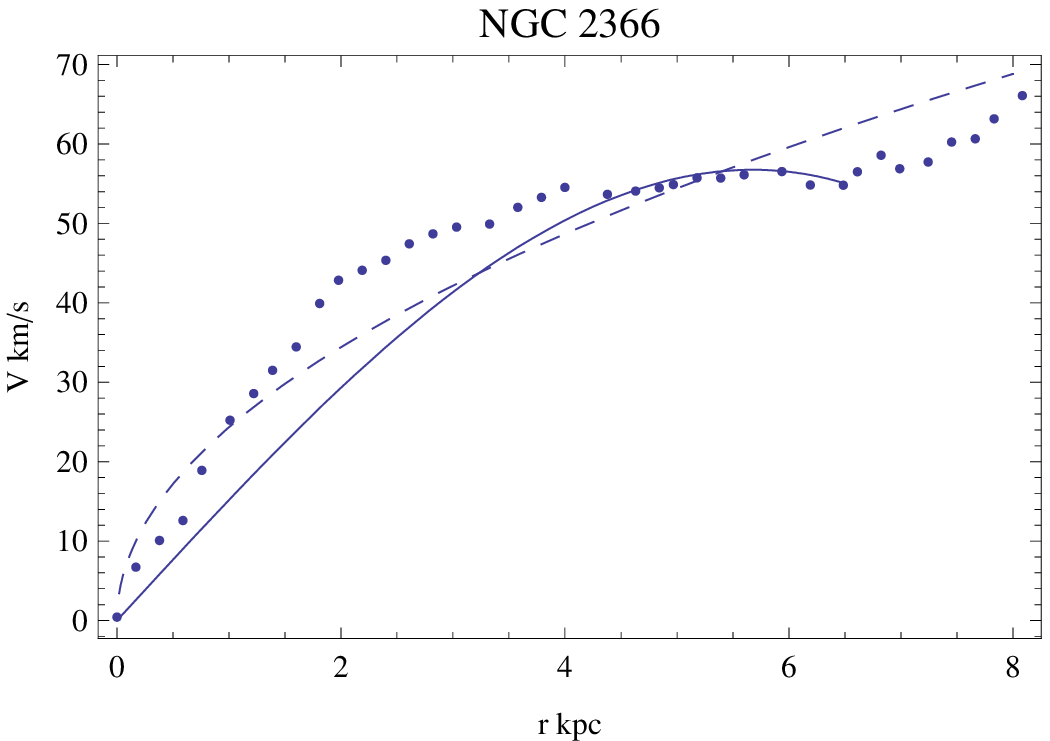}\\[1ex]
\includegraphics[width=0.48\linewidth]{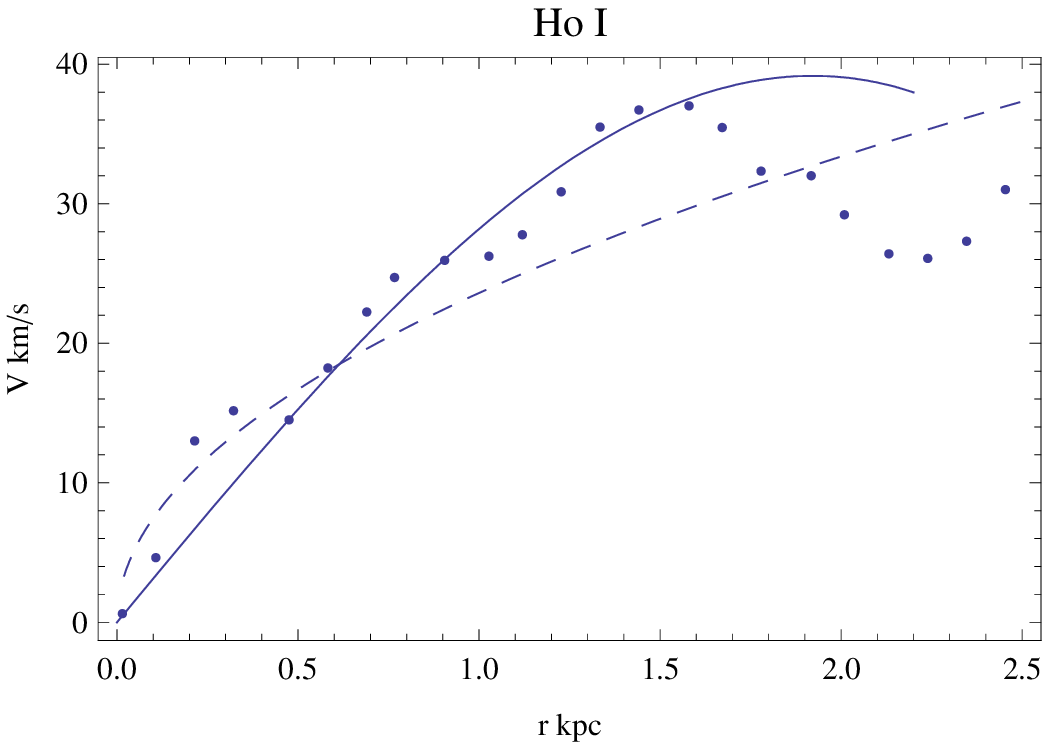}
\includegraphics[width=0.48\linewidth]{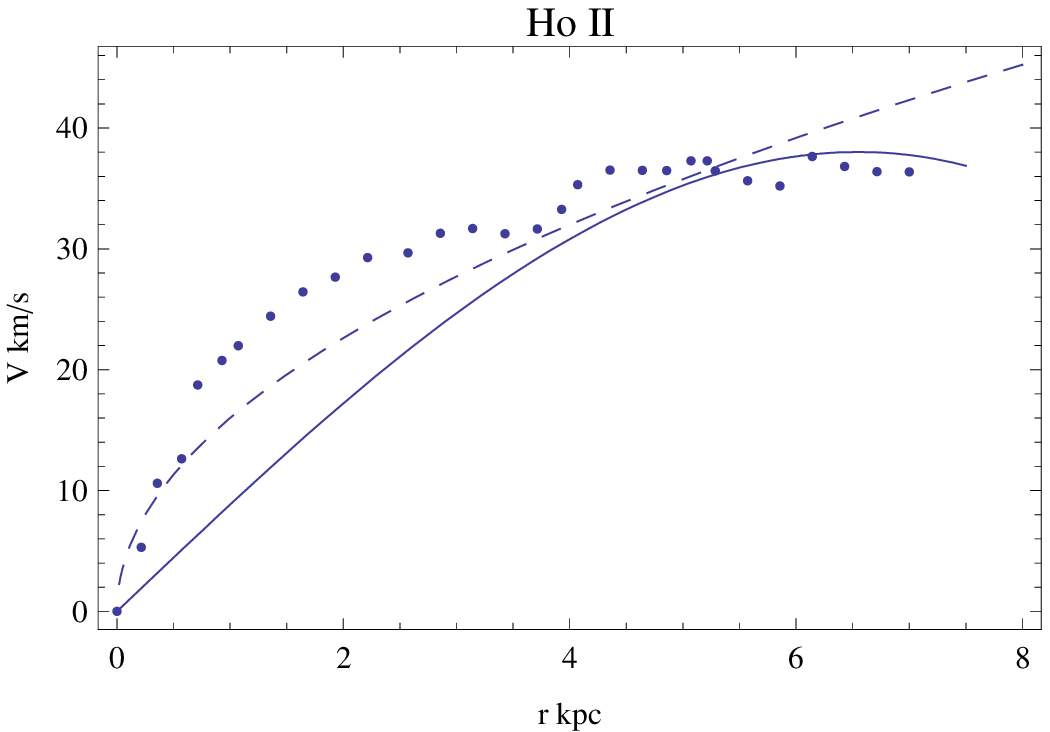}\\[1ex]
\includegraphics[width=0.48\linewidth]{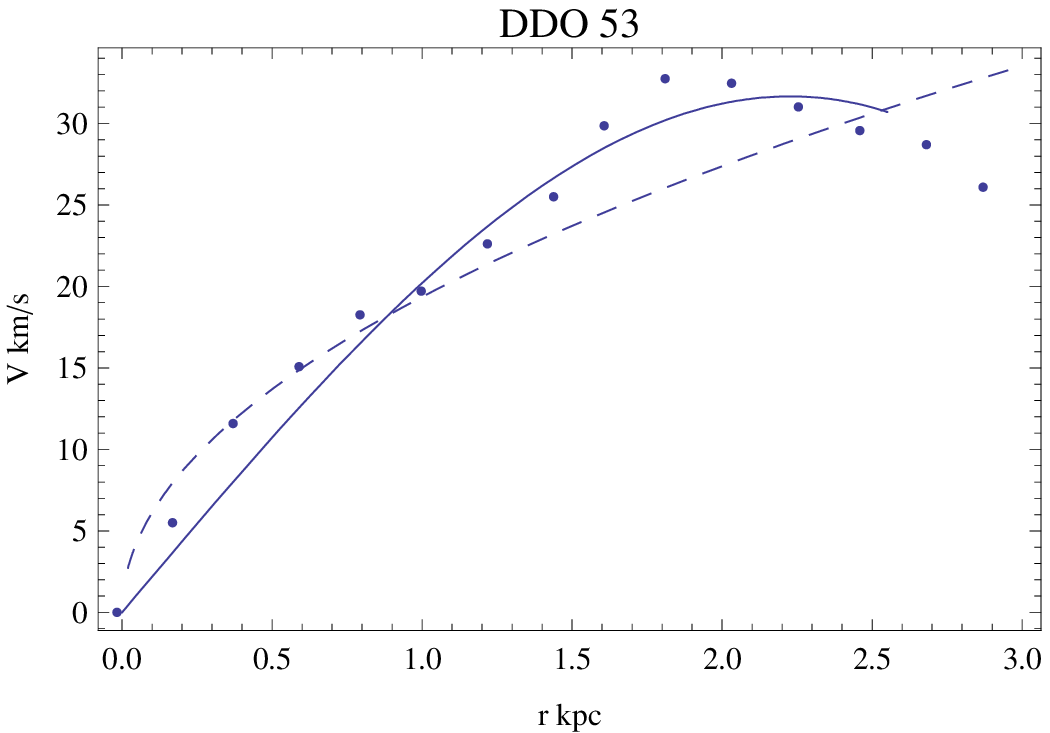}
\includegraphics[width=0.48\linewidth]{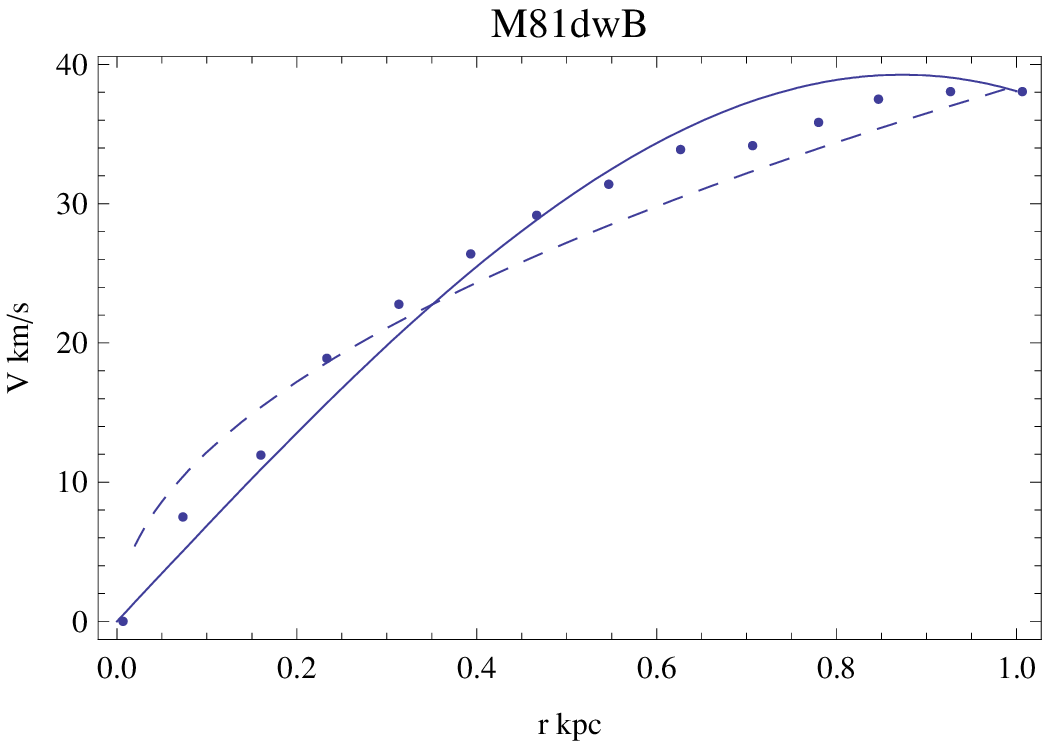}\\[1ex]]
\includegraphics[width=0.48\linewidth]{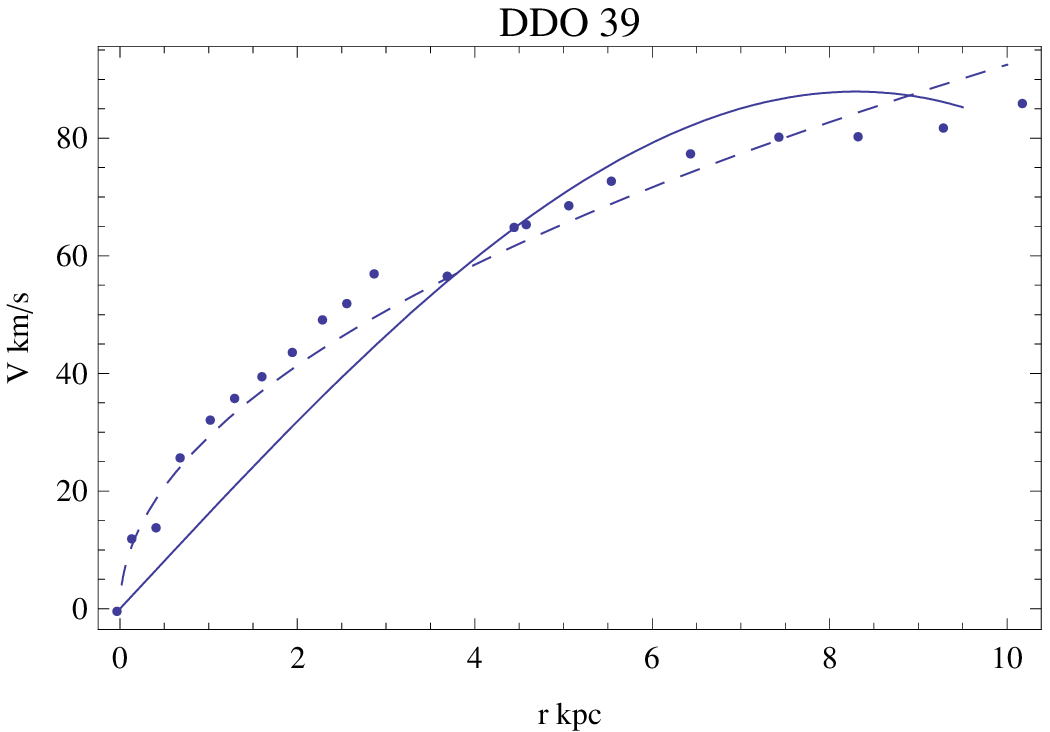}
\includegraphics[width=0.48\linewidth]{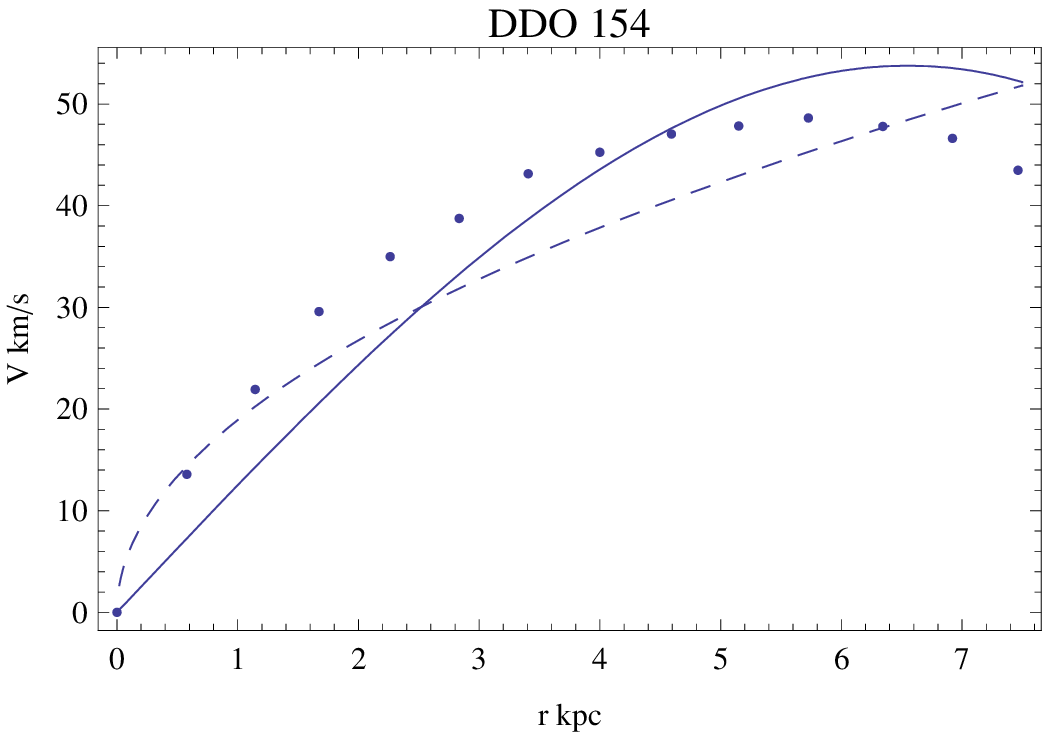}\\[1ex]
\caption{Comparison of the velocity of the rotation curves predicted by the Bose-Einstein Condensate dark matter model (solid curve), of the velocity fits obtained with the NFW density profile (dashed curve), and of the observed rotational curves of eight dwarf galaxies, respectively.} \label{fig1}
\end{figure}

The rotation curves predicted by the BEC model and the NFW profile, respectively, are presented,
together with the observational data,  in Fig.~\ref{fig1}. The numerical values of the two free
parameters of the BEC model (central density and radius) are shown in Table~\ref{table1}.

\begin{widetext}
\begin{table}
\begin{center}
\begin{tabular}{|c|c|c|c|c|c|c|c|c|}
\hline
Galaxy & IC 2574 & NGC 2366 & Ho I & Ho II & DDO 53 & M81 dwB & DDO 39 & DDO
154 \\ \hline
$R$ (kpc) & 12.6 & 6.5 & 2.2 & 7.0 & 2.54 \  & 1.0 & 9.5 & 7.5 \\ \hline
$\rho _{c}$ $\left( 10^{-24}\;\mathrm{g/cm}^{3}\right) $ & 0.44 & 0.89 & 3.7
& 0.35 & 1.8 & 18.0 & 1.04 & 0.60 \\ \hline
\end{tabular}
\caption{Radii and central densities of the dwarf galaxies used in the predicted BEC
rotation curves.}
\label{table1}
\end{center}
\end{table}
\end{widetext}

The parameters obtained by fitting the NFW model with the observational results are presented in Table~\ref{table2}.

\begin{widetext}
\begin{table}
\begin{center}
\begin{tabular}{|c|c|c|c|c|c|c|c|c|}
\hline
Galaxy & IC 2574 & NGC 2366 & Ho I & Ho II & DDO 53 & M81 dwB & DDO 39 & DDO
154 \\ \hline
$\rho _{s}$ ($\;{\rm g/cm}^{3}$) & 25272.5 & 48553.7 & 45877.9 & 32341.2 &
38835.4 \  & 73763.8 & 57633.0 & 37276.1 \\ \hline
$r_{s}$ $\left( {\rm km}\right) $ & 49407.1 & 29113.2 & 27598.2 & 18885.3 &
3030.7 & 47856.5 & 35439.0 & 22927.1 \\ \hline
\end{tabular}
\caption{The numerical values of the NFW parameters $\rho _s$ and $r_s$ obtained by fitting the NFW rotational velocity profile.}
\label{table2}
\end{center}
\end{table}
\end{widetext}

As one can see from Fig.~\ref{fig1},  there is an overall  good agreement
between the observed data and the theoretical predictions of the BEC model.  The good agreement with the observations strongly suggest that our model may be relevant for obtaining a correct description of the dark matter, and its properties. In all predictions and fits the effect of the baryonic matter was completely neglected. Even that baryonic matter may represent only 10\% of the entire mass of the dwarf galaxy, by including its gravitational effects, one can obtain a significantly better description of the rotation curves, especially in the small $r$ regions. One should also mention that the model gives a relatively good description of the rotation curve of the galaxy DDO 154, which is generally difficult to be achieved. However, the considered cases shows a variety of behaviors.
Below we discuss the results obtained  for the rotation curves for each galaxy.

$\bullet$ IC 2574: For this galaxy the BEC model gives an almost perfect prediction for the rotation curve. The NFW velocity also gives a good fit. However, is is not as good as the BEC model prediction. The galaxy may be affected by non-circular motions \cite{Oh}, but the rotation velocity, with a maximum of around  80 km/s is the highest in the sample. The galaxy is also the most massive in the considered sample, with $M=1.462\times 10^{10}M_{\odot}$ \cite{Oh}. With the use of the BEC model mass - radius relation, $M=\left(4/\pi \right)\rho _cR^3$, we obtain for the mass of the galaxy the value $M_{ICS574}^{BEC}=1.64\times 10^{10}M_{\odot}$, which is very close to the observed value.

$\bullet $ NGC 2366: The rotation curve of this galaxy is rather complicated, indicating that disturbances caused by non-circular motions could be present in the outer regions with $r> 5$ kpc. Non-rotational motions may also be present for $r<5$ kpc, distorting the inner region of the rotation curve. It is difficult to give an estimate of the radius of the dark matter distribution, because of the increasing trend in the velocity field for $r$ between 6 and 8 kpc. The adopted values for the radius and central density give an estimate of the mass of the galaxy as $M_{NGC2365}^{BEC}=4.54\times 10^{9}M_{\odot}$, while the observed mass is $M=4.29\times 10^9M_{\odot}$ \cite{Oh}. The BEC velocity profile gives a good description of the rotation curve for $r\in(4.5,6.5)$ kpc. The NFW fit gives a reasonable good description of the rotation curve.

$\bullet $ Holmberg I: For $r>1.5$ kpc, the rotation curve of the galaxy Holmberg I shows a severe distortion, probably caused by the presence of strong external gravitational interactions, generating non-circular motions. Therefore the physical parameters and behavior of the galaxy without the presence of external perturbation are difficult to be extracted from the observed data.  With the adopted values for the radius and central density, $R=2.2$ kpc and $\rho _c=3.7\times 10^{-24}$ g/cm$^3$, the mass of the galaxy is obtained as $M_{HOI}^{BEC}=0.76\times 10^{9}M_{\odot}$ (observed mass $M=0.46\times 10^9M_{\odot}$ \cite{Oh}). However, the BEC model gives a very good description of the inner region $r<1.5$ kpc of the rotation curve. The NFW velocity fit fails to give any reasonable description of the rotation curve.

$\bullet $ Holmberg II: Rather large velocity dispersions, as compared to what would be expected for a circular rotation velocity only, appear in the rotation curve of the dwarf galaxy Holmberg II.   Non-circular motions and strong baryonic gravitational perturbations in the galaxy are the most probably explanation for the pattern of the rotation curve for $r<5$ kpc.  With the values of the radius and central density adopted to predict the rotation curve the mass of the galaxy is obtained as $M_{HOII}^{BEC}=2.76\times 10^{9}M_{\odot}$ (observed mass $M=2.07\times 10^9M_{\odot}$ \cite{Oh}). The BEC model describes well the outer ($r>5$ kpc) region of the galaxy, but fails to describe the inner galactic regions, where any realistic model must take into account the presence of supplementary (local) interactions. The NFW fit cannot reproduce consistently the observed data.

$\bullet $ DDO 53: The rotation curve of the galaxy DDO 53 shows a clear and smooth rotational pattern, even that the possibility of external perturbations for $r>2$ kpc cannot be excluded. These perturbations may be responsible for the sudden decay of the curve, without the presence of a significant plateau. However, the BEC prediction describes well the rotation curve. The mass of the galaxy is obtained as  $M_{HOII}^{BEC}=0.55\times 10^{9}M_{\odot}$, while the observed mass is $M=0.45\times 10^9M_{\odot}$ \cite{Oh}. The NFW profile describes well the inner region of the rotation curves, but fails to give a good description of the curve in the outer galactic regions.

$\bullet $ M81 dwB: The BEC model gives a very good description of the rotation curve of the galaxy. This galaxy is the smallest in the sample, having a mass of only $M=0.3\times 10^9M_{\odot}$ \cite{Oh}. With the adopted values for radius and central density the mass of the galaxy is $M_{M81dwB}^{BEC}=0.33\times 10^{9}M_{\odot}$. The velocity field shows no significant non-circular motions in the galaxy, except perhaps in the very outer regions. The NFW fit cannot reproduce the main features of this rotation curve.

$\bullet $ DDO 39: The rotation curve of the galaxy DDO 39 shows the presence of significant non-circular motions in the inner galactic region with $r<1.8$ kpc. A strong gravitational perturbation may be responsible for the distortion of the inner part of the rotation curve. Other non-circular motions may be present in the outer region of the curve, for $r>9$ kpc. The BEC model gives a relatively good description of the observational data in the middle and outer regions, where the motion is mostly rotational. The NFW fit gives a good description of the rotation curve. The adopted radius and central density for the BEC model gives a mass of $M_{DDO39}^{BEC}=1.6\times 10^{10}M_{\odot}$.

$\bullet $ DDO 154: The rotational velocity curve of the galaxy DDO 154 shows a regular rotation pattern. No significant non-circular motions can be observed in the galaxy. The rotation curve resembles that of a galaxy with solid-body rotation, with a steep increase in the inner regions. The BEC rotation curve is obtained by taking $R=7.5$ kpc and $\rho _c=6\times 10^{-25}$ g/cm$^3$, which gives a mass of $M_{DDO154}^{BEC}=4.71\times 10^{9}M_{\odot}$. The mass of the galaxy determined from observations is $M=5.40\times 10^9M_{\odot}$ \cite{Oh}. The BEC prediction describes reasonably well the outer region of the rotation curve. The NFW profile cannot give a satisfactory fit to the observational data.

\subsection{Density profiles and mean logarithmic density and velocity slopes}

\begin{figure}[tbh]
\centering
\includegraphics[width=0.78\linewidth]{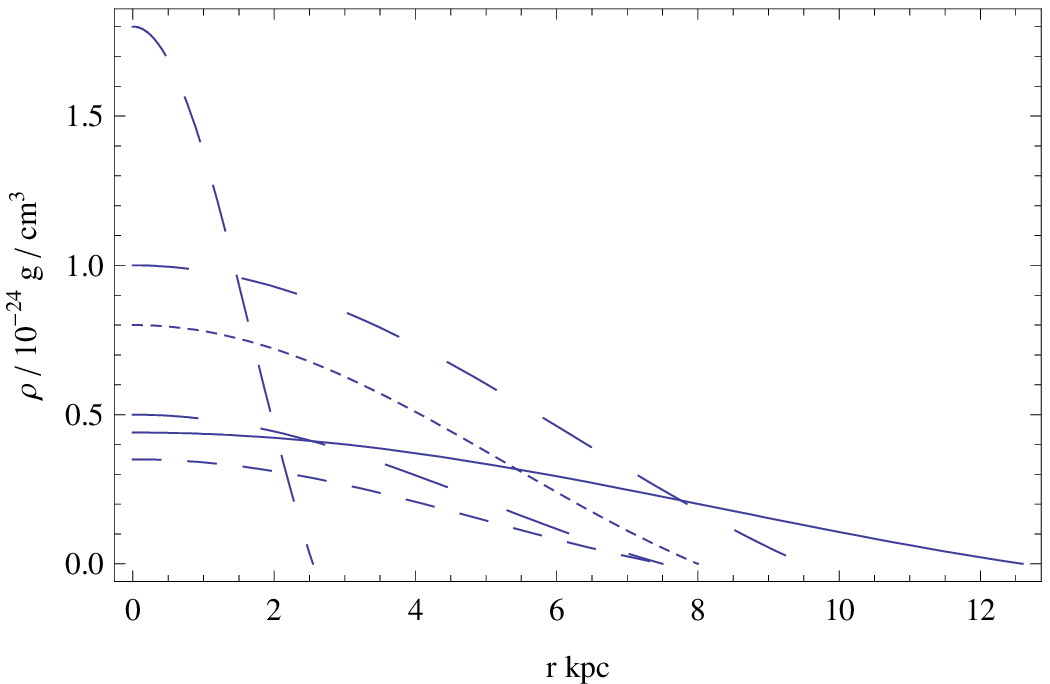}
\caption{Density profiles of the galaxies IC 2574 (solid curve), NGC 2366 (dotted curve), Ho II (short dashed curve), DDO  53 (dashed curve), DDO 39 (long dashed curve) and DDO 154 (ultra-long dashed curve), respectively.} \label{fig2}
\end{figure}

\begin{figure}[tbh]
\centering
\includegraphics[width=0.78\linewidth]{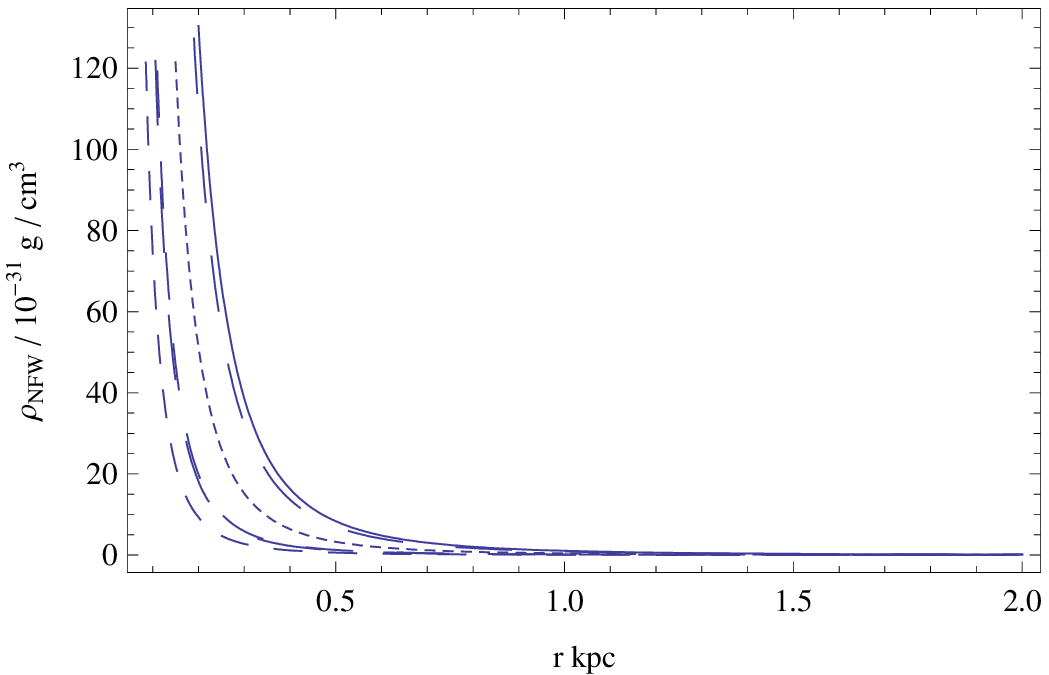}
\caption{NFW density profiles of the galaxies IC 2574 (solid curve), NGC 2366 (dotted curve), Ho II (short dashed curve), DDO  53 (dashed curve), DDO 39 (long dashed curve) and DDO 154 (ultra-long dashed curve), respectively.} \label{fig3}
\end{figure}

The corresponding density profiles of the dark matter halos are presented in
Fig.~\ref{fig2} for the BEC model, while the NFW density profiles are shown in Fig.~\ref{fig3}. The density profile of the BEC condensate is non-singular, and well behaved at the center of the galaxy. Depending on the numerical values of the radius and central density, the mass distribution of the gravitationally bounded condensate decreases slowly as a function of $r$, with most of the matter concentrated in a core-like region. This theoretical prediction of the BEC model is in sharp contrast with the cusp-like structure found in the numerical simulations \cite{nfw}, which can be seen in Fig.~\ref{fig3}. The NFW density profile is singular at the center of the galaxy, for $r\rightarrow 0$, and it increases very rapidly for small $r$. On the other hand, for large $r$ the NFW profile tends more rapidly to zero than the BEC profile. Another important difference between the two profiles is that the BEC profile predicts a finite and well defined radius for the dark matter distribution, whose density tends rigourously to zero on a surface that define the radius $R$ of the dark matter halo. For the NFW density profile a well-defined radius does not exists. Therefore Bose-Einstein condensation processes
in dwarf galaxies can affect the dark matter distribution in such a way that the central cusps predicted from dark-
matter-only cosmological simulations are flattened, resulting in dark matter halos characterized by a large and  almost constant density core, as observationally found in normal dwarf galaxies in the local universe \cite{Wa,Oh}.

As a next step in our analysis we consider the behavior of the logarithmic
density profile slope $\alpha $, defined as
\begin{equation}
\alpha =\frac{d\log \rho }{d\log r},
\end{equation}
and which can be easily expressed as
\begin{equation}
\alpha (r)=-\left[ 1-kr\cot (kr)\right]
=-\frac{\pi ^{2}V^{2}(r)}{4\pi G\rho R^{2}}.
\end{equation}

For $r\rightarrow R$, $\alpha (r)$ diverges, $\alpha \left( R\right) \rightarrow \infty $. The mean value of the
logarithmic inner slopes of the mass density profiles have been obtained
from the observations of the rotations curves of dwarf galaxies \cite{Oh} and
low surface brightness galaxies \cite{deBl}, respectively. We define the mean
value of the inner $\alpha $ as
\begin{equation}
\left\langle \alpha _{in}\right\rangle
=\frac{ 1}{R_{in}}\int_{0}^{R_{in}}\alpha
(r)dr=-\int_{0}^{x_{in}}\left( 1-\pi x\cot \pi x\right) dx,
\end{equation}
where $R_{in}$
is the radius of the inner core of the dark matter distribution, and $%
x_{in}=R_{in}/R$.

\begin{figure}[tbh]
\centering
\includegraphics[width=0.78\linewidth]{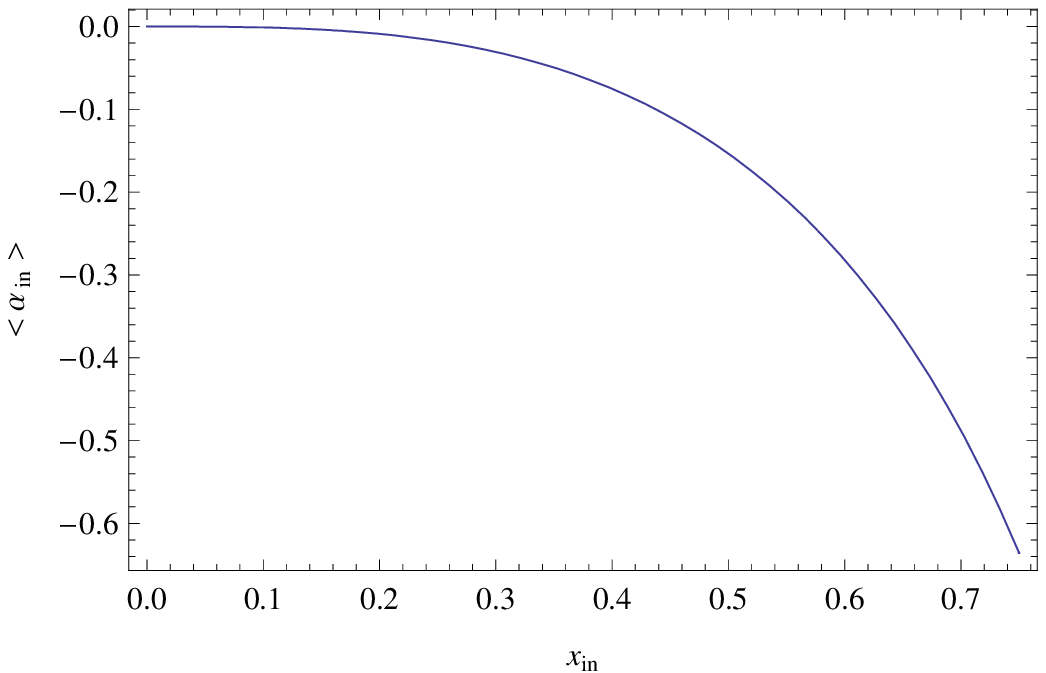}
\caption{The mean value of the
logarithmic inner slope of the mass density profile of dwarf galaxies.} \label{fig4}
\end{figure}

The variation of $\left\langle \alpha _{in}\right\rangle $
as a function of $x_{in}$ is represented in Fig.~\ref{fig4}.

For $R_{in}=0.6R$ we
obtain $\left\langle \alpha _{in}\right\rangle =-0.2818$, a value which is
very close to the value $-0.29\pm 0.07$ obtained in \cite{Oh}. Therefore in the BEC dark matter model a value of $\left\langle \alpha _{in}\right\rangle $ in agreement with the observations can be easily obtained. Of course,
the procedure proposed in the present paper for the determination of $\left\langle \alpha
_{in}\right\rangle $ is very different from the method used in \cite{Oh},
where first a break-radius, where the slope changes most rapidly, was
determined. Then the inner density slope was measured by performing a least
squares fit to the observational data points within the break-radius. In the theoretical estimation of $\left\langle \alpha _{in}\right\rangle$ we have assumed that $R_{in}=0.6R$ is the break-radius. In the CDM model the value of $\alpha $ is $\alpha =-1$.

The
logarithmic slope of the rotation curves is defined as
\begin{equation}
\beta =\frac{d\log V}{d\log r},
\end{equation}
 and in the Bose-Einstein condensate dark halo model it can be obtained
as
\begin{eqnarray}
\beta (r)&=&-\frac{1}{2}\left[ 1-\frac{k^{2}r^{2}}{ 1-kr\cot (kr)} \right]=\nonumber\\
&& -\frac{1}{2}\left( 1-\frac{4\pi G\rho r^{2}}{V^{2}(r)}\right) .
\end{eqnarray}
Therefore for Bose-Einstein condensed dark matter halos the logarithmic
velocity and density slopes are related by the simple relation
\begin{equation}
\beta (r)=-\frac{1}{2}\left[ 1+\frac{\pi ^{2}}{\alpha (r)}\left( \frac{r}{R}%
\right) ^{2}\right] .
\end{equation}
This theoretical prediction represents another possibility for a direct observational test of the condensed dark matter model.

\section{Discussions and final remarks}\label{4}

The Bose-Einstein condensation takes place when particles (bosons) become
correlated with each other. This happens when their wavelengths overlap,
that is, the thermal wavelength $\lambda _{T}=\sqrt{2\pi \hbar ^{2}/mk_{B}T}$
is greater than the mean inter-particles distance $a$, $\lambda _{T}>a$. The
critical temperature for the condensation to take place is $T_{cr}<2\pi \hbar
^{2}n^{2/3}/mk_{B}$ \cite{Da99}. On the other hand, cosmic evolution has the same temperature dependence, since
in an adiabatic expansion process the density of a matter dominated Universe evolves as $\rho \propto T^{3/2}$ \cite{Fuk09}. Therefore, if the boson temperature is equal, for example, to the radiation temperature at $z = 1000$,
 the critical temperature for the Bose-Einstein condensation is at present $T_{cr} = 0.0027K$ \cite{Fuk09}. Since the matter temperature $T_m$ varies as $T_m\propto  a^{-2}$, where $a$ is the scale factor of the Universe, it follows that during an adiabatic evolution the ratio of the photon temperature $T_{\gamma }$ and of the matter temperature evolves as $T_{\gamma }/T_m\propto a$. Using for the present day energy density of the Universe
the value $\rho _{cr}= 9.44 \times 10^{-30}$ g/cm$^3$, BEC takes place provided that the boson mass satisfies the restriction $m < 1.87$ eV \cite{bec1}.
Thus, once the
temperature $T_{cr}$ of the boson is less than the critical temperature,  BEC can always take place at some
moment during the cosmological evolution of the Universe. On the other hand,  we expect that the Universe is always under critical temperature, if it is at the present time \cite{Fuk09}.

Another cosmological bound on the mass of the condensate particle can be obtained as $m<2.696\left(g_d/g\right)\left(T_d/T_{cr}\right)^3$ eV \cite{Bo}, where $g$ is the number of internal degrees of freedom of the particle before decoupling, $g_d$ is the number of internal degrees of freedom of the particle at the decoupling, and $T_d$ is the decoupling temperature. In the Bose condensed case $T_d/T_c < 1$, and it follows that the BEC particle should be light, unless it decouples very early on, at high temperature and with a large $g_d$. Therefore,  depending on the relation between the critical  and the decoupling
temperatures, in order for a BEC light relic to act as cold dark matter,  the decoupling scale must be higher than the electroweak scale \cite{Bo}.

In the Bose-Einstein condensate dark matter model the physical properties of the galaxies are determined by three parameters: the scattering length $a$, the mass of the particle $m$, which completely determine the radius of the dark matter distribution $R=R(a,m)$, and the central density of the dark matter, which is necessary for the estimation of the mass $M=M\left(a,m,\rho _c\right)$. In both these quantities $a$ and $m$ enter in the combination $a/m^3$. If the radius of the galactic dark matter distribution is known, then the ratio of the two intrinsic parameter that determine the properties of the Bose Einstein condensed dark matter halos can be obtained as
\begin{equation}
\left( \frac{a}{{\rm fm}}\right) \left( \frac{m}{{\rm eV}}\right) ^{-3}=3278.38\left(
\frac{R}{{\rm kpc}}\right) ^{2}.
\end{equation}

\begin{figure}[tbh]
\centering
\includegraphics[width=0.78\linewidth]{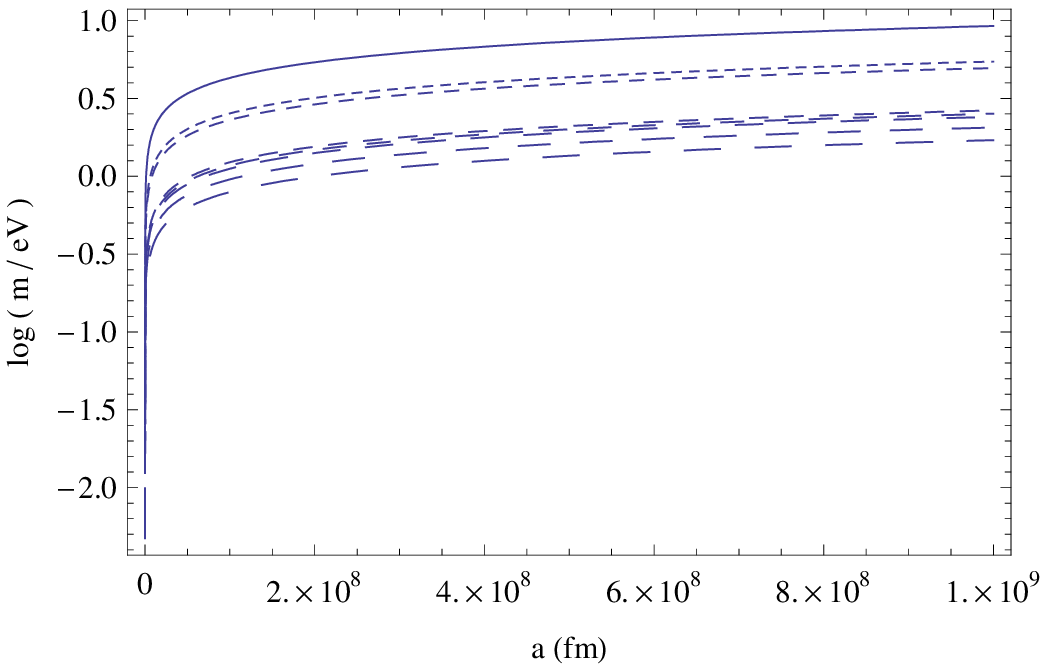}
\caption{Mass (in a logarithmic scale) of the dark matter particle as a function of the scattering length $a$ for the considered sample of galaxies, in increasing radius order: M81 dwB (solid curve), Ho I, DDO 53, NGC 2366, DDO 154, DDO 39, and IC 2574, respectively.  } \label{fig5}
\end{figure}

For the considered sample of dwarf galaxies the ratio $a/m^3$ varies between $3278.38$ fm/eV$^3$ (obtained from the galaxy M81 dwB with radius $R=1$ kpc) and $504,084$ fm/eV$^3$ for the galaxy IC 2574, with radius $R=12.4$ kpc. Since the variation of the radius for the considered sample of galaxies is around one order of magnitude, the range of $a/m^3$ obtained from individual galaxies can differ by two orders of magnitude. By taking the mean value of the radius $R_{mean}=6.5$ kpc, we obtain $a/m^3=122189$ fm/eV$^3$. A very small particle mass, of the order of $10^{-22}$ eV would require a scattering length of the order of $a\approx 1.22\times 10^{-61}$ fm, which seems to be too small to be realistic. The variation of the mass of the dark matter particle as a function of the scattering length $a$ is represented, for each considered galaxy, in Fig.~\ref{fig5}.

\begin{figure}[tbh]
\centering
\includegraphics[width=0.78\linewidth]{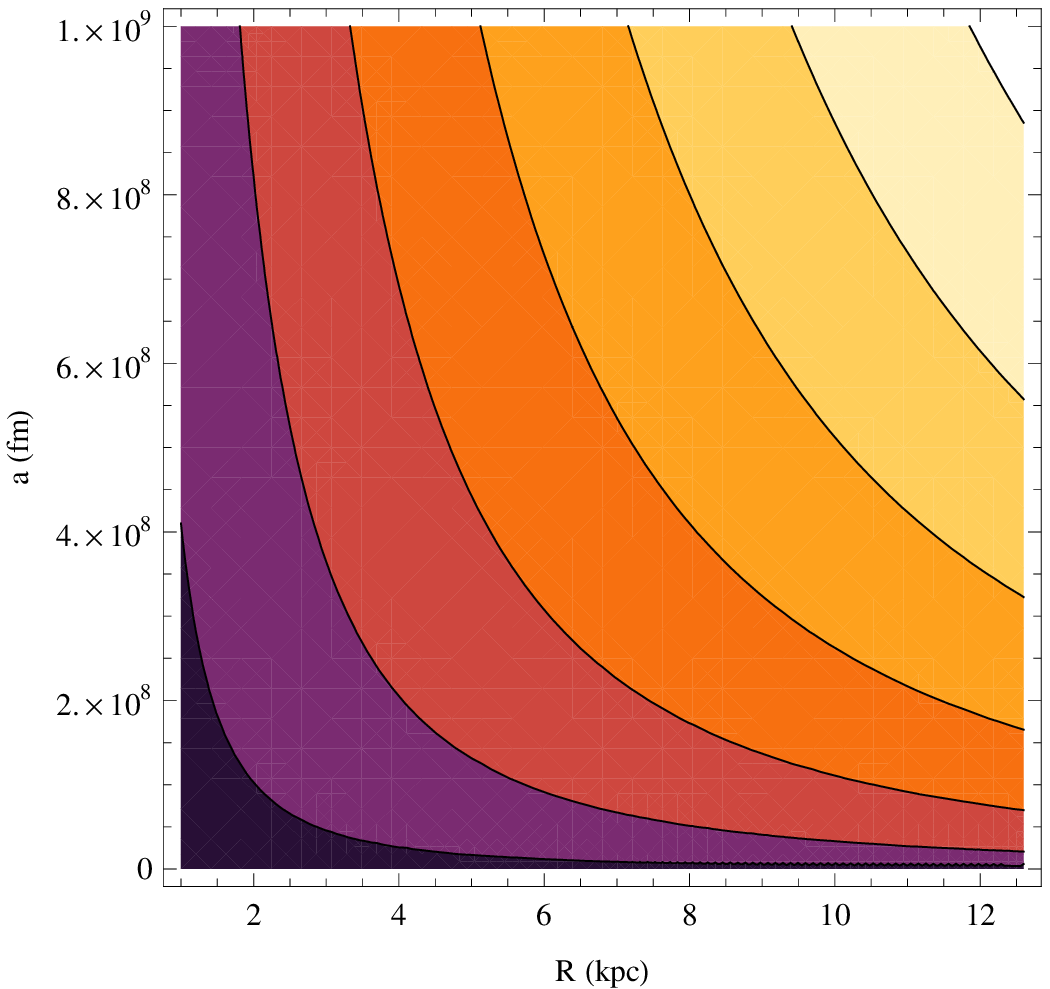}
\caption{Contour plot of the mass $m$ of the dark matter particle as a function of the scattering length $a$ and of the radius $R$ of the dwarf galaxy.} \label{fig6}
\end{figure}

The contour plot of the mass of the dark matter particle as a function of the galactic radius $R$ and of the scattering length $a$, $m=m\left(a,R\right)$ is represented in Fig.~\ref{fig6}. A contour plot, which is an alternative to a three dimensional surface plot, is a graphical technique for representing a three-dimensional surface by plotting constant $m$ slices, called contours, on a two-dimensional format. That is, for a given value of $m$, the curves shown in the figure connect the points with coordinates $\left(R,a\right)$  where that $m$ value occurs. In order to consistently derive the values of $a$ and $m$ from observations another relation between these two parameters is necessary. Such a relation could be obtained from cosmological considerations, as suggested in \cite{Hu}.

Finally, we would like to address the problem of the mass of the dwarf galaxies. As compared to the standard estimation of the mass $M\approx 4\pi R^3\rho _c/3$, the total mass of the Bose-Einstein condensate, given by Eq.~(\ref{mass}), is smaller by a factor of three for constant density halos, and can be one order of magnitude smaller for halos with a non-constant density profile. This may solve the mass discrepancy between observations and simulations, which was pointed out in \cite{Ti}. As compared to the BEC models,
the usually core-like models, such as pseudo-isothermal halo models, dominated by
a central constant-density core \cite{Oh}, predict a much higher galaxy mass.

Since dwarf galaxies are usually considered to be dark matter dominated, the possibility of reproducing their rotation curves, and the mean value of the logarithmic inner slope of the mass density profiles, represents a significant evidence in favor of the strongly - coupled dilute Bose-Einstein condensate dark matter model. Moreover, the density profile of the condensate has generally a large core with a slowly decreasing density. Another important advantage of the model is that all physical quantities are well behaved for all $r$, and there are no singularities in any of the physical parameters.
In the present paper we have shown that the strongly correlated Bose-Einstein  condensate dark matter model gives generally a relatively good description of the properties of the dark matter dominated dwarf galaxies. All the relevant physical quantities can be predicted from the model, and can be directly compared with the corresponding observational parameters (the dark halo mass, the radius of the galaxy, and the observed flat rotational velocity curves, mean values of the density profiles etc.). Therefore
this opens the possibility of in depth testing of the Bose-Einstein condensation models by using astronomical and astrophysical observations at the galactic-intergalactic scale. In this paper we have provided some basic theoretical tools necessary for the comparison of the predictions of the condensate model and of the astronomical results.

\section*{Acknowledgments}

I would like to thank to the anonymous referee for comments and suggestion that helped me to significantly improve the manuscript. This work was supported by a GRF grant of the government of the Hong Kong SAR.


\begin{thebibliography}{99}

\bibitem{PeRa03}  P. J. E. Peebles and B. Ratra, Rev. Mod. Phys. \textbf{75}%
, 559 (2003); T. Padmanabhan, Phys. Repts. \textbf{380}, 235 (2003).

\bibitem{nfw}  J. F. Navarro, C. S. Frenk, and S. D. M. White, Astrophys. J.
\textbf{490}, 493 (1997).

\bibitem{bur}  A. Burkert, Astrophys. J. Letters \textbf{447}, L25 (1995).

\bibitem{Wa} F. Walter, E. Brinks, W. J. G. de Blok, F. Bigiel, R. C.  Kennicutt,
M. Thornley, and A. Leroy,  Astron. J. {\bf 136}, 2563 (2008).

\bibitem{Oh} S.-H. Oh, W. J. G. de Blok, E. Brinks, F. Walter, and R. C. Kennicutt,  Jr., arXiv:1011.0899 (2010).

\bibitem{deBl} W. J. G. de Blok, S. S. McGaugh, A. Bosma, and V. C. Rubin,  Astrophys. J. {\bf 552}, 23 (2001).

\bibitem{Ti} T. Sawala, Q. Guo, C. Scannapieco, A. Jenkins and S. White, arXiv:1003.0671 (2010).

\bibitem{Lo} A. Loeb and N. Weiner, arXiv:1011.6374 (2010).

\bibitem{Oh1} S.-H. Oh, C. Brook, F. Governato, E. Brinks, L. Mayer, W. J. G. de Blok, A. Brooks, and F. Walter, 	 arXiv:1011.2777 (2010).

\bibitem{Sin} S. J. Sin, Phys. Rev. D{\bf 50}, 3650 (1994);  S. U. Ji and S. J. Sin, Phys.
Rev. D{\bf 50}, 3655 (1994).

\bibitem{fer} F. Ferrer and J. A. Grifols, JCAP {\bf 0412}, 012 (2004).

\bibitem{Fu05} T. Fukuyama and M. Morikawa, Progress of Theoretical Physics
\textbf{115}, 1047 (2006).

\bibitem{fer1} J. A. Grifols, Astropart. Phys. {\bf 25}, 98 (2006).

\bibitem{BoHa07}  C. G. Boehmer and T. Harko, JCAP \textbf{06}, 025 (2007).

\bibitem{exp}  M. H. Anderson, J. R. Ensher, M. R. Matthews, C. E. Wieman
and E. A. Cornell, Science \textbf{269}, 198 (1995); C. C. Bradley, C. A.
Sackett, J. J. Tollett and R. G. Hulet, \prl {\bf 75}, 1687 (1995); K. B.
Davis, M. O. Mewes, M. R. Andrews, N. J. van Drutten, D. S. Durfee, D. M.
Kurn and W. Ketterle, \prl {\bf 75}, 3969 (1995).

\bibitem{bec1} T. Fukuyama, M. Morikawa, and T. Tatekawa, JCAP {\bf 0806}, 033 (2008).

\bibitem{bec2}J.-W. Lee, Phys. Lett. B {\bf 681}, 118 (2009).

 \bibitem{bec3} M. N. Brook and P. Coles, arXiv:0902.0605 (2009).

\bibitem{Fuk09} T. Fukuyama and M. Morikawa, Phys. Rev. D {\bf 80}, 063520 (2009).

\bibitem{Ri} T. Rindler-Daller and P. R. Shapiro, Vortices and Angular Momentum in Bose-Einstein-Condensed Cold Dark Matter Halos, ASP Conerence Series {\bf 432},  244; arXiv:0912.2897 (2010).

\bibitem{bec4} B. Kain and H. Y. Ling, Phys. Rev. D {\bf 82}, 064042 (2010).

\bibitem{LEE} J.-W. Lee and S. Lim, JCAP {\bf 01}, 007 (2010).

\bibitem{Bon} A. P. Lundgren, M. Bondarescu, R. Bondarescu, and J. Balakrishna, Astrophys. J. Lett. {\bf 715}, L35 (2010).

\bibitem{bec5} T. Harko, arXiv:1101.3655, accepted for publication in MNRAS, (2011).

\bibitem{Ch} P.-H. Chavanis, arXiv:1103.2050 (2011); P.-H. Chavanis and L. Delfini, arXiv:1103.2054 (2011); P.-H. Chavanis, arXiv:1103.2698 (2011); P.-H. Chavanis, arXiv:1103.3219 (2011).

\bibitem{Ma} T. Matos and A. Suarez, arXiv:1103.5731 (2011).

\bibitem{Guz} J. A. Gonzalez and F. S. Guzman, Phys. Rev. D {\bf 83}, 103513 (2011). 

\bibitem{Hu} W. Hu, R. Barkana, and A. Gruzinov, Phys. Rev. Lett. {\bf 85},  1158 (2000).

\bibitem{Da99}  F. Dalfovo, S. Giorgini, L. P. Pitaevskii and S. Stringari,
Rev. Mod. Phys. \textbf{71}, 463 (1999); C. J. Pethick and H. Smith,
	Bose-Einstein condensation in dilute gases, Cambridge, Cambridge University Press, (2008).

\bibitem{Bo}  D. Boyanovsky, H. J. de Vega, and N. Sanchez, Phys. Rev. D
\textbf{77}, 043518 (2008).

\bibitem{Sw} R. A. Swaters, M. A. W. Verheijen, M. A. Bershady, and D. R. Andersen, 	Astrophys. J. {\bf 587}, L19 (2003).

\bibitem{Ca} C. Carignan and S. Beaulieu, Astrophys. J. {\bf  347}, 760 (1989).

\end{thebibliography}
\end{document}